\documentclass[twocolumn,prl,aps,showpacs]{revtex4}
\usepackage{amssymb}
\usepackage{amsfonts}
\usepackage{amsmath}
\usepackage{epsfig}
\usepackage{graphicx}

\begin{document}

\title{

{Accuracy of the semiclassical picture of photoionization in intense laser fields}
\footnotetext{$^{\dag}$ sshan@mail.shcnc.ac.cn}
\footnotetext{$^{\ddag}$ chen$\_$jing@iapcm.ac.cn}
 }

\author{Li Guo$^{1}$}
\author{Shilin Hu$^{2}$}
\author{Mingqing Liu$^{3,4}$}
\author{Zheng Shu$^{3,4}$}
\author{XiWang Liu$^{5}$}
\author{Jie Li$^{5}$}
\author{Weifeng Yang$^{5}$}
\author{Ronghua Lu$^{1}$}
\author{ShenSheng Han$^{1,{\dag}}$}
\author{Jing Chen$^{2,3,{\ddag}}$}

\affiliation{$^{1}$ Key laboratory for Quantum Optics and Center for Cold
Atom Physics, Shanghai Institute of Optics and Fine Mechanics,
Chinese Academy of Sciences, Shanghai 201800, China}
\affiliation{$^{2}$ Laboratory of Quantum Engineering and Quantum Metrology,
School of Physics and Astronomy, Sun Yat-Sen University (Zhuhai
Campus), Zhuhai 519082, China}
\affiliation{$^{3}$ HEDPS, Center for Applied Physics and Technology, Collaborative Innovation Center of IFSA,
Peking University, Beijing 100084, China}
\affiliation{$^{4}$ Institute of Applied Physics and Computational Mathematics, P. O.
Box 8009, Beijing 100088, China}
\affiliation{$^{5}$
Department of Physics,
College of Science, Shantou University, Shantou, Guangdong 515063,
China}


\begin{abstract}
In the semiclassical picture of photoionization process in intense
laser fields, the ionization rate solely depends on the amplitude
of the electric field and the final photoelectron momentum
corresponds to the instant of ionization of the photoelectron,
however, this picture has never been checked rigorously. Recently
an attosecond angular streaking technique based on this
semiclassical perspective has been widely applied to temporal
measurement of the atomic and molecular dynamics in intense laser
fields. We use a Wigner-distribution-like function to calculate
the time-emission angle distribution, angular distribution and
ionization time distribution for atomic ionization process in
elliptically polarized few-cycle laser fields. By comparing with
semiclassical calculations, we find that the two methods always
show discrepancies except in some specific cases and the offset
angles are generally not consistent with the offset times of
the ionization time distributions obtained by the two methods even
when the non-adiabatic effect is taken into account, indicating
that the ``attoclock" technique is in principle inaccurate.
Moreover, calculations for linearly polarized laser fields also show
similar discrepancies between two methods in the ionization time
distribution. Our analysis indicates that the discrepancy between
the semiclassical and quantum calculations can be attributed to
correlation, i. e., temporal nonlocalization effect.

\end{abstract}

 \maketitle
\section{Introduction}
Study on above-threshold ionization (ATI) plays an essential role
in our comprehension of various phenomena in strong-field atomic
physics \cite{ap79,wbecker1}. So far, a categorization of the
ionization process has been widely accepted: two distinct regimes-
multiphoton ionization and tunneling ionization are distinguished
by the Keldysh parameter $\gamma=\sqrt{\frac{I_{p}}{2U_{p}}}$ with
$\gamma\ll1$ for tunneling ionization and $\gamma\gg1$ for
multiphoton ionization \cite{keldysh}. Here, $I_{p}$ is the
ionization potential of the atom and $U_{p}$ is the ponderomotive
energy. In the multiphoton regime, the ionization occurs via
absorption of $n$ photons from the field. In the tunneling regime,
the electron is considered to tunnel through the barrier created
by the superposition of the laser field and Coulomb potential and
thereafter can be treated as a free electron evolving in
the external laser field. This is commonly known as the
simpleman's picture of the ATI process \cite{h}, which comprises
the picture to understand atomic and molecular dynamics in intense
laser fields \cite{wbecker1,Corkum1} and constitutes the
foundation of attosecond measurement \cite{KrauszRMP2009}.

In this semiclassical picture, the ionization rate of the electron
is only determined by the instantaneous amplitude of the laser
field and the final photoelectron momentum is equal to the vector
potential of the laser field at the ionization moment with
opposite sign if the influence of the ionic potential is ignored,
however, this correspondence, or the accuracy of the semiclassical
approximation of the photoionization process, has never been
rigorously checked, or in other words, how accurate the
semiclassical picture is still remains an open question. Recently,
an attosecond angular streaking technique, also dubbed as
"attoclock" technique, has been developed to investigate the
temporal dynamics of atoms and molecules in intense laser fields
\cite{Eckle1,Eckle2,pfeiffer2011,Eckle3,boge,wu2012PRL,optica}.
This technique, different from the conventional attosecond
measurement which relies on attosecond pulses generated using
high-order harmonic generation process and is very technically
demanding, uses the rotating electric-field vector of an intense
close-to-circularly polarized pulse to deflect photo-ionized
electrons in the radial spatial direction. Then the instant of
ionization is mapped to the final angle of the momentum vector in
the polarization plane according to the semiclassical picture.
When a few-cycle pulse is applied, a comparison between the peaks of
the measured photoelectron angular distribution and the
simpleman's prediction shows an offset angle which has caused much
debate on its underlying mechanism
\cite{Eckle2,Eckle3,boge,Iva,optica}. Ionic Coulomb potential,
tunneling time delay and nonadiabatic effect have been proposed to
explain this offset angle. By comparing the experimental data with
the semiclassical calculation including the Coulomb potential,
Eckle \emph{et al}. place an intensity-averaged upper limit of 12
attoseconds on the tunneling delay time in strong field ionization
of a helium atom \cite{Eckle2}. Boge \emph{et al}. find that the
nonadiabatic effect is unimportant \cite{boge}. However, by
solving a three-dimensional time-dependent Schr\"{o}dinger
equation, Ivanov \emph{et al}. \cite{Iva} give an opposite view
against the calculations using the semiclassical model
\cite{boge}. Recently, Torlina \emph{et al}. show that the
offset angle can be attributed to the Coulomb potential effect and
confirm the zero tunneling time delay by theoretical calculation
\cite{Torlina2015}.

It is noteworthy that the principle of the attosecond streaking
technique, viz., the angle of the final momentum vector
corresponds to the instant of the tunneling ionization, is based
on the semiclassical picture. Therefore, prior to consideration of
the Coulomb potential and nonadiabatic effects \emph{etc}.
\cite{Eckle2,Yudin2001,Barth2011,Wang2014,pfe,hofmann,ni,hofmann2,km},
one needs to rigorously check the principle of the attoclock
technique, or the accuracy of the semiclassical picture in
description of the photoionization process which is an intrinsic
quantum process.

In this paper, we use a Wigner-distribution-like (WDL) function
\cite{oe-guo,pra-guo,jpb-guo} to investigate the ATI process of an
atom in few-cycle laser pulses in the framework of the
strong-field approximation (SFA). It is worthwhile mentioning that
the SFA is accurate for a short-range system, e. g., negative ion,
so it is suitable to investigate the validity of the semiclassical
picture or the principle of the original attoclock technique
scheme in which the Coulomb potential is ignored in the first
place. The WDL function enables us to calculate the time-emission
angle distribution, angular distribution and ionization time
distribution of the ATI process of atom in elliptically and
circularly polarized laser pulses with different carrier-envelop
phases (CEPs). Then we explicitly and rigorously check the
validity and accuracy of the attosecond angular streaking
technique by comparing the quantum distributions with results
given by the semiclassical simulation. Moreover, we also calculate
the time-energy distribution and ionization time distribution of
atoms in linearly polarized laser pulses and compare with
semiclassical simulations to further systematically check the
accuracy of the semiclassical picture in description of the
photoionizaiton process. Furthermore, we consider the
non-adiabatic effect by performing calculation with PPT theory.
The result shows that the non-adiabatic effect indeed reduces the
discrepancy between semiclassical and quantum distributions and
improves the accuracy of the attoclock technique.

\section{Theory and numerical methods}
The Wigner-distribution-like
function derived from the strong-field approximation model is
defined as (see \cite{oe-guo} for more details):
\begin{equation}
f(t,\frac{p^{2}}{2})=\frac{1}{\pi}\int_{-\infty}^{\infty}S'^{*}(t+t')S'(t-t')e^{-2i\frac{p^{2}}{2}t'}dt'.\label{1}
\end{equation}
where $S'$ is given by
\begin{align}
S'   =&\frac{\sqrt{2\pi}}{\sqrt{v}}\frac{\partial\phi_i({\textbf{q}})}{\partial{\textbf{q}}
} \cdot\mathbf{E }(t)\nonumber \\
&\times\exp\left\{i\int_{-\infty}^{t}[\mathbf{p}\cdot
\mathbf{A}(\tau)+\frac{\mathbf{A}^2(\tau)}{2}]d\tau+iI_pt\right\}.\label{re}
\end{align}
Here, $v$ is normalization volume, $\mathbf{E }(t)$ the electric
field and $I_p$ the ionization potential of the atom.
$\phi_{i}(\textbf{q})$ is the Fourier transform of
 the atomic ground state $\vert\varphi
_{0}\rangle$ and ${\textbf{q}}\equiv {\mathbf{p}}+{\mathbf{A}}(t)$.

In this paper, we consider a laser pulse with a sin$^{2}$-type
temporal envelop. The vector potential
is given by
\begin{align}
\textbf{A}(t)=&-\frac{E_{0}}{\omega}\sin^{2}\left[\frac{\omega
t}{n}\right]\nonumber\\ &
\left[\cos\frac{\theta}{2}\cos(\omega t+\varphi)\textbf{e}_{x}\right.
\left.-\sin\frac{\theta}{2}\sin(\omega
t+\varphi)\textbf{e}_{y}\right],\label{3}
\end{align}
where $E_{0}$ is the peak field strength, $\omega$ the laser
frequency, and n/2 the number of optical cycles contained in the
laser pulse. $\varphi$ is the initial carrier envelop phase (CEP).
$\textbf{e}_{x}$ and $\textbf{e}_{y}$ are the unit vectors along the
x and y axes, respectively. $\epsilon=\cot\frac{\theta}{2}$ is the
ellipticity. The major axis is y-axis.

For simplification, here we only consider the distribution
of electrons emitted in the polarization plane for a
three-dimensional system. For a fixed kinetic energy
$\frac{p^2}{2}$ in the polarization plane, the electron emission angle $\Theta$ varies from
0 to 2$\pi$, where $\Theta$ is the angle between the final
momentum of electrons and x axis. From Eq. (\ref{re}), we can find
that $S'$ is a function of momentum $\mathbf{p}$, which can be
written as a function of two scalar variables $p$ and $\Theta$ in
a two-dimensional model. Here $p$ is the absolute value of the
electron momentum vector. Therefore, the formula of the
Wigner-distribution-like function for a two-dimensional system can
be written as following:

\begin{equation}
f(t,\frac{p^{2}}{2},\Theta)=\frac{1}{\pi}\int_{-\infty}^{\infty}S'^{*}(t+t',\Theta)S'(t-t',\Theta)e^{-2i\frac{p^{2}}{2}t'}dt'.\label{4}
\end{equation}
The time-angle distribution function is given by
\begin{equation}\label{5}
 f'(t,\Theta)=\int f(t,\frac{p^{2}}{2},\Theta)d(\frac{p^2}{2}).
\end{equation}

Then one can find that the ionization probability as functions of
time and emission angle can be given by
\begin{equation}
P(t)=\int f'(t,\Theta)d\Theta. \label{7}
\end{equation}
and
\begin{equation}\label{8}
 W(\Theta)=\int f'(t,\Theta)dt.
\end{equation}

\section{Results and discussions}
In this paper, we consider H atoms ionized by a 6-cycle elliptically
polarized laser field with peak intensity of $1\times10^{14}$
W/cm$^2$ and ellipticity $\epsilon=0.882$. It is worthwhile
mentioning that all integrations are performed numerically in our
quantum calculations.

\begin{figure}[tbh]
\begin{center}
\includegraphics[width=8cm,height=4.5cm]{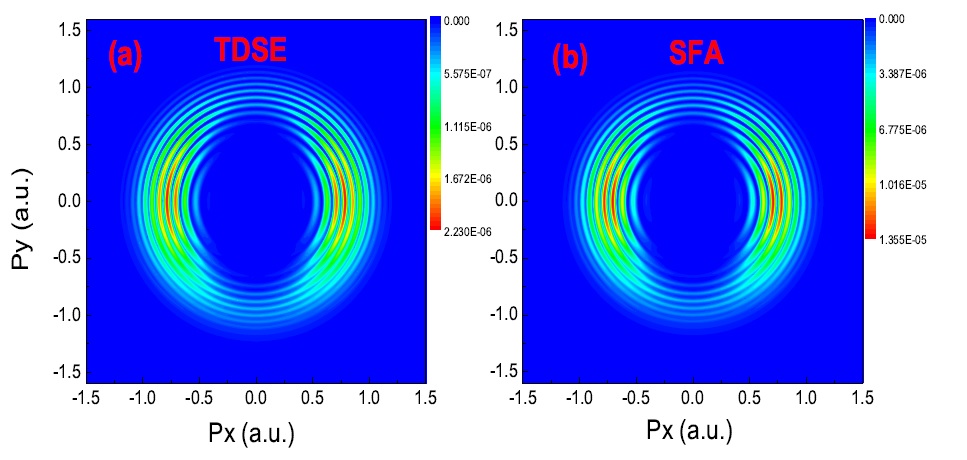}
\end{center}
\caption{Momentum spectra calculated by TDSE (a) and SFA (b) methods.
}%
\label{fig01}%
\end{figure}

At first, we show the comparison of momentum spectra given
by numerical solution of time-dependent
Schr$\mathbf{\ddot{o}}$dinger equation (TDSE)(for details, see
\cite{tdse1,tdse2}) and the SFA theory with the laser parameters of
the CEP $\varphi=0.5\pi$ and $\omega$=0.05691 a.u. in Fig.
\ref{fig01}. The short-range potential of $V=-A\frac{e^{-\kappa
r}}{r}$ is adopted with $\kappa=1$ in the TDSE method, where the value of $A$
is taken as 1.90847 in order to make the model potential have the ground sate energy of -0.5 a.u.. The momentum spectra in Fig. \ref{fig01} show distinct ATI rings and are in good agreement with each
other.

\subsection{Elliptical and circular polarization}
 The semiclassical method in this paper is
based on the ADK model \cite{Brabec1996PRA,Hu14,Hao11} without
taking into account the ionic Coulomb potential, which corresponds
to the SFA model considered in our WDL calculation. The ionization rate in the semiclassical result is obtained
using Eq. (21) of \cite{krainov} with the Coulomb correction factor removed. The
distribution of the initial momentum at the tunnel exit is
$P(p_{0\perp},p_{0\parallel})\propto$
exp$(-\frac{p_{0\perp}^{2}}{|E(t)|})$exp$(-\frac{p_{0\parallel}^{2}}{|E(t)|})$
($E(t)$ is the laser field amplitude at time $t$. $p_{0\parallel}$
and $p_{0\perp}$ are the initial momenta parallel and vertical
to the direction of the laser field polarization, respectively)
\cite{Hao11}. When only initial transverse momentum is considered,
we have $p_{0\parallel}=0$.  For clarity, we label the two
different semiclassical methods according to the different initial
momentum adopted as ADK$_{\perp}$ with only the initial transverse
momentum considered and ADK$_{\perp,\parallel}$ with both the
initial transverse and longitudinal momenta considered.

\begin{figure}[tbh]
\begin{center}
\includegraphics[width=8.cm,height=8.2cm]{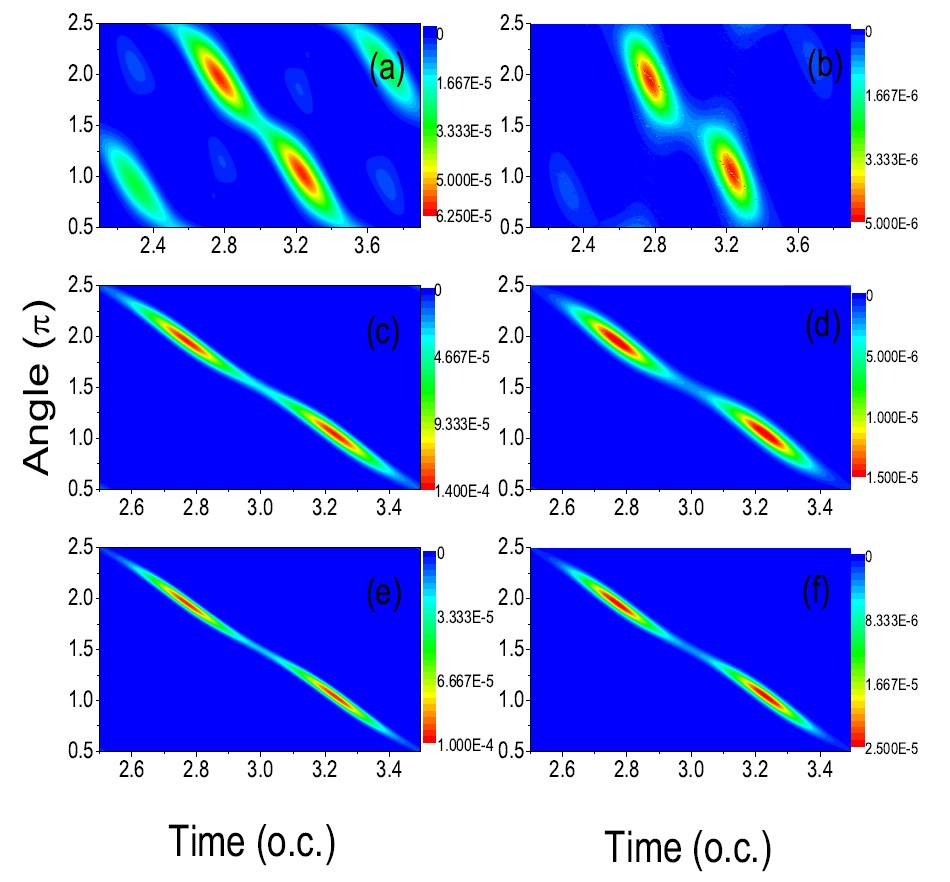}
\end{center}
\caption{ Time-emission angle distributions in a
6-cycle laser pulse with the CEP $\varphi=0.5\pi$ for different
laser frequencies $\omega=$0.182 a.u. ((a)
and (b)), 0.05691 a.u. ((c) and (d)) and 0.03502 a.u. ((e) and
(f)). Peak intensity $I=1\times10^{14}$ W/cm$^2$ and the
ellipticity $\epsilon$=0.882. Quantum results:
 (a), (c), and (e) ; calculations of the semiclassical theory with ADK$_{\perp,\parallel}$: (b), (d), and (f).
}%
\label{fig1}%
\end{figure}

Figure \ref{fig1} shows the calculation of the time-emission angle
distribution for H atoms in a 6-cycle laser pulse with the CEP
$\varphi=0.5\pi$ for three different optical frequencies. The
distributions in Fig. \ref{fig1} are obtained by Eq. (\ref{5})(see
Figs. \ref{fig1}(a), \ref{fig1}(c), and \ref{fig1}(e))
and by semiclassical theory with ADK$_{\perp,\parallel}$ (see
Figs. \ref{fig1}(b), \ref{fig1}(d), and \ref{fig1}(f)), respectively.
For the semiclassical calculation obtained by ADK$_{\perp,\parallel}$, all distributions with
different laser frequencies look similar- two main peaks located
at [$\Theta, t$]$\sim$ [$0(2\pi)$, 2.75 o.c.] and [$\pi$, 3.25
o.c.] (o.c. is the abbreviation of optical cycle)
and two additional small peaks at $t\sim$ 2.3 and 3.7 o.c. (not
shown in Fig. \ref{fig1}(d) and \ref{fig1}(f)) which correspond to
the secondary peaks of the field amplitude. This is expectable
since the ionization is independent of the frequency in the
quasistatic tunneling picture. In contrast, for the quantum
calculation, the time-emission angle distributions show a clear
transition with decreasing the laser frequency.
For $\omega$=0.182 a.u. ($\gamma$=3.4), the distribution
(Fig. \ref{fig1}(a)) is similar to the structure of the
semiclassical calculation (Fig. \ref{fig1}(b)). When the laser
frequency further decreases to $\omega$=0.05691 a.u.
($\gamma$=1.07) and 0.035 a.u. ($\gamma$=0.66), the quantum
results more and more mimic the semiclassical results. Therefore,
Fig. \ref{fig1} demonstrates a transition from the multiphoton
regime to the tunneling regime in the elliptically polarized laser
field, which is consistent with the case of linearly polarized
laser fields \cite{pra-guo}. It is worthwhile mentioning
that the time-emission angle distribution is also calculated using
ADK$_{\perp}$ method, however, the width is much narrower than the
quantum result and is hardly dependent on the laser frequency (not
shown here), indicating that the initial longitudinal momentum is
an important non-adiabatic effect even in the tunneling regime.

\begin{figure}[tbh]
\begin{center}
\includegraphics[width=8.cm,height=8.cm]{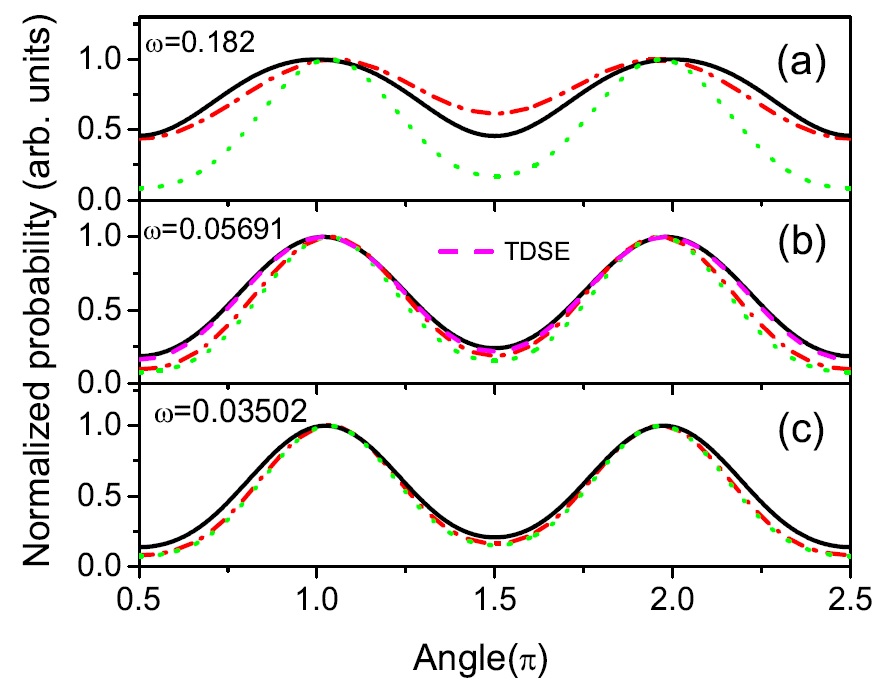}
\end{center}
\caption{Angular distributions calculated by the quantum theory
(solid lines) and semiclassical theories with
ADK$_\perp,_\parallel$ (dashed dotted lines) and ADK$_\perp$
(dotted lines) for different laser frequencies with the same
parameters as those in Fig. \ref{fig1}. The angular
distribution given by the TDSE for $\omega$=0.05691 a.u. is also
present for comparison.
}%
\label{fig2}%
\end{figure}

To investigate quantitatively the accuracy of the semiclassical
theory and validity of the attoclock technique, we then calculate
the angular distribution $W(\Theta)$ by integrating the
time-emission angle distribution over time $t$ \cite{note} and
compare the quantum calculations with the semiclassical ones (see
Fig. \ref{fig2}). All the angular distributions calculated by the
semiclassical model ADK$_{\perp,\parallel}$ (dashed dotted lines) show a double-peak
structure, which is symmetrical with respect to the angel
$\Theta$=1.5$\pi$, in accordance with those shown in
Figs. \ref{fig1}(b), \ref{fig1}(d), and \ref{fig1}(f). However, the widths of the angular distributions
become progressively broader with the increasing frequency. We
also give the results obtained by ADK$_{\perp}$ (dotted lines),
which also display a symmetric double-peak. The distinct feature
different from the distribution gained by ADK$_{\perp,\parallel}$
is that the width of the angular distribution hardly depends on
the frequency. By comparing between these two kinds of ADK
calculations, we can infer that the initial longitudinal momentum
distribution can cause broader width of the angular distribution,
resulting in the difference between these two ADK calculations. This difference
becomes larger when the frequency increases.

Similar to Fig. \ref{fig1}, the quantum calculation also clearly
shows a transition from the multiphoton regime to the tunneling
regime with decreasing laser frequency comparing with the
semiclassical calculation.
For $\omega$=0.182 a.u., the angular distribution in the quantum calculation
shows a double-peak pattern with peak positions close to the
semiclassical distribution obtained by ADK$_{\perp,\parallel}$ (see Fig. \ref{fig2}(a)). When the frequency decreases further,
the quantum distribution becomes gradually close to the
semiclassical calculations (see Figs. \ref{fig2}(b) and (c)).
However, the widths of the quantum distributions are wider than
the semiclassical ones and positions of peaks in the quantum and
semiclassical calculations are still noticeable different (Here,
we define the difference between the peak positions in the angular
distributions calculated by the quantum theory and other methods
(e.g. ADK models and non-adiabatic calculation mentioned below) as
offset angle denoted by $\triangle \Theta$). The angular distribution gained by TDSE is also given in Fig. \ref{fig2}(b), which almost coincides with the curve of SFA model. The reason that the curves of the angular distributions calculated by TDSE and SFA models do not completely overlap is that the potential used in the TDSE calculation still possesses a finite range.
Moreover, the effect
of the initial longitudinal momentum of the photoelectron, which
is considered as the non-adiabatic effect in the tunneling
ionization process \cite{pfe,nt,cn} can also be clearly seen in
Fig. \ref{fig2}. For the lowest frequency $\omega$=0.03502 a.u.,
the two different semiclassical calculations (ADK$_{\perp}$ and
ADK$_{\perp,\parallel}$) can hardly be distinguished. For
$\omega$=0.05691 a.u., the distribution of ADK$_{\perp,\parallel}$
becomes wider than that of ADK$_{\perp}$ and is closer to the
quantum distribution. The difference between two semiclassical
calculations becomes larger when the frequency increases to
$\omega$=0.182 a.u., however, the distribution of
ADK$_{\perp,\parallel}$ becomes even wider than the quantum
distribution. This is understandable since it is already in the
multiphoton regime ($\gamma$=3.4), the non-adiabatic effect cannot
be treated properly by the semiclassical model.

\begin{figure}[tbh]
\begin{center}
\includegraphics[width=8.cm,height=6.cm]{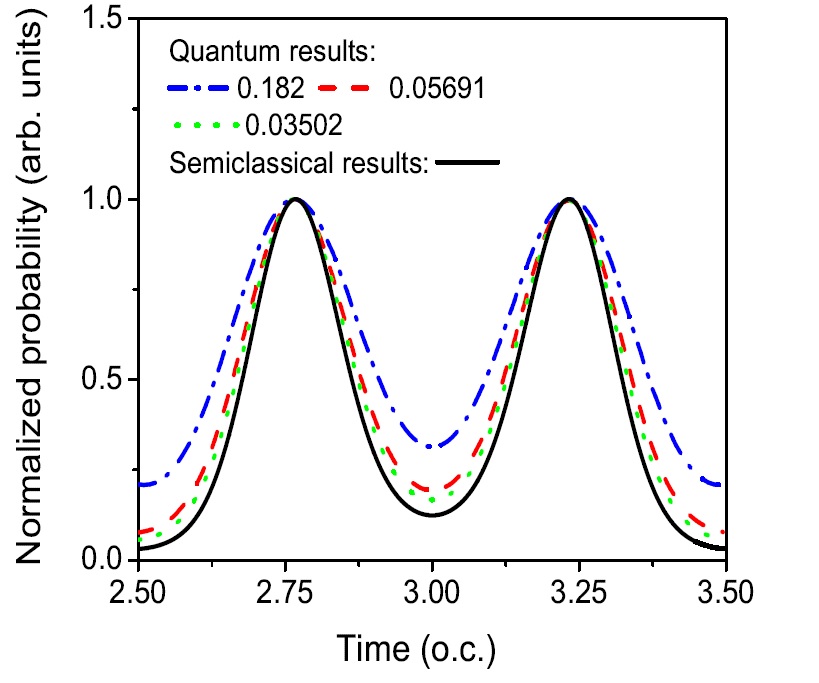}
\end{center}
\caption{ Ionization time distributions calculated
by the quantum theory and semiclassical theories with
ADK$_\perp,_\parallel$ and ADK$_\perp$ (solid line) for different
laser frequencies with the same parameters as those in Fig.
\ref{fig1}.
}%
\label{fig3}%
\end{figure}

\begin{figure}[tbh]
\begin{center}
\includegraphics[width=8.5cm,height=8.5cm]{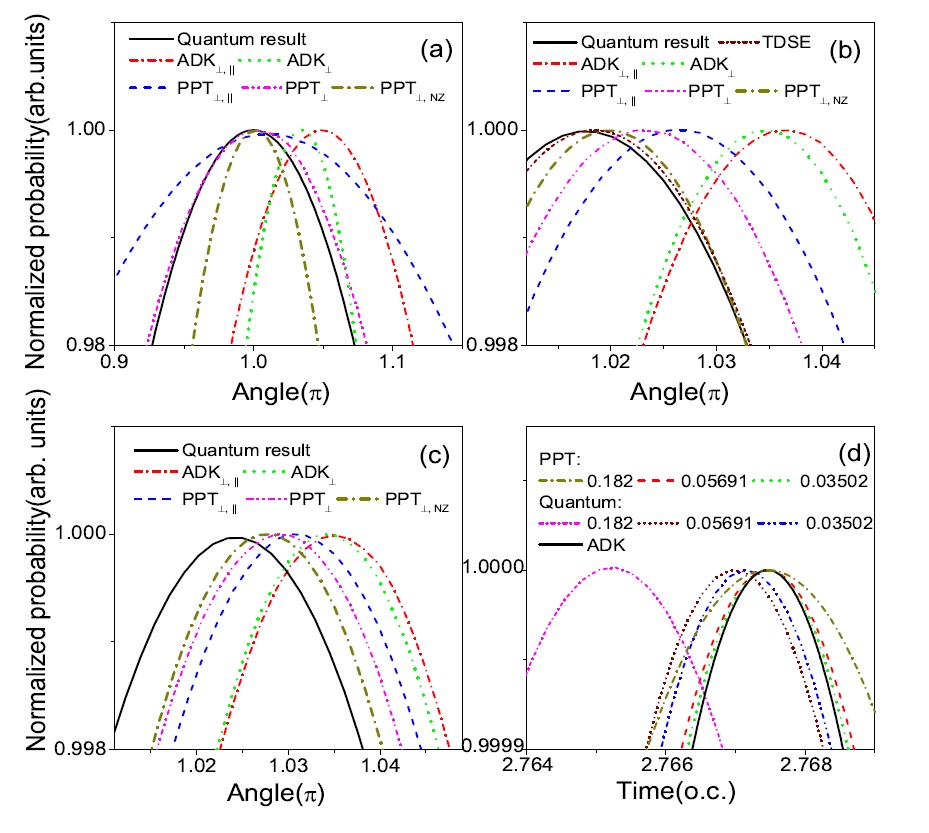}
\end{center}
\caption{ Angular distributions calculated by the
quantum, semiclassical and PPT theories for different laser
frequencies $\omega=0.182$ a.u.(a), $\omega=0.05691$ a.u. (b) and
$\omega=0.03502$ a.u. (c); (d): ionization time distributions
calculated by different methods (see text).
}%
\label{fig4}%
\end{figure}

Moreover, we calculate the ionization time distribution $P(t)$ by
integrating the time-emission angle distribution over angle
$\Theta$ and the results are depicted in Fig. \ref{fig3}. For the
semiclassical simulation, the ionization time distributions
obtained by both ADK$_{\perp,\parallel}$ and ADK$_{\perp}$ are the
same since the initial momentum distributions in two calculations
are both normalized. Furthermore, it is expected that the
ionization probability is only dependent on the electric field
strength but independent of the laser frequency. As shown in Fig.
\ref{fig3}, the ionization time distributions obtained by the
semiclassical methods (black solid line) exhibit a double-peak
structure for the time range from $t=2.5$ to 3 o.c. and are indeed
independent of the laser frequency, whose peak positions
correspond to the maxima of the electric field strength
($t$=2.76746 and 3.23254 o.c.). For the quantum calculation, the
distribution varies with the frequency.
All ionization time distributions possess a double-peak structure
and the widths of the peaks decrease with decreasing laser
frequencies, becoming more and more close to the semiclassical
results. However, the difference between the positions of peaks in
the ionization time distributions calculated by the quantum method
and semiclassical model still exists (for more details, see Fig.
\ref{fig4}). We define this difference in the ionization time
distribution as offset time denoted by $\bigtriangleup t$.

It should be noted that, for such wide frequency regime
considered here which covers from tunneling to multiphoton regime,
a theory including non-adiabatic effect should be used for
calculations of both the angular and ionization time distributions.
These calculations will be shown in Fig. \ref{fig4} but here we
only show the results of the adiabatic theory for comparison with
the quantum result to show clearly approach of the quantum result
to the semiclassical calculation with decreasing the frequency.

In order to show the offset angle and offset time more clearly, we
zoom in one of two peaks in the angular and ionization time
distributions. Figs. \ref{fig4}(a)-\ref{fig4}(c)
show the angular distributions for $\omega$=0.182
a.u., 0.0569 a.u. and 0.03502 a.u., respectively. As shown in
Figs. \ref{fig4}(a)-\ref{fig4}(c), it can be clearly seen that the
widths of the angular distributions calculated by
ADK$_{\perp,\parallel}$ (red dashed dotted lines) are broader than
those gained by ADK$_{\perp}$ (green dotted lines) and the peak
position for ADK$_{\perp},_{\parallel}$ is a little farther from
the quantum peak position comparing with ADK$_{\perp}$
calculation. These differences between two ADK calculations result
from the initial longitudinal momentum as mentioned above. When
the frequency decreases, these differences also decrease, which
can be attributed to relatively larger momentum of the
electron acquired from the laser field with lower frequency
comparing with the wavelength-independent initial
momentum of the electron at the tunnel exit
\cite{Brabec1996PRA,Hu14,Hao11}.
\begin{figure}[tbh]
\begin{center}
\includegraphics[width=7.cm,height=7.cm]{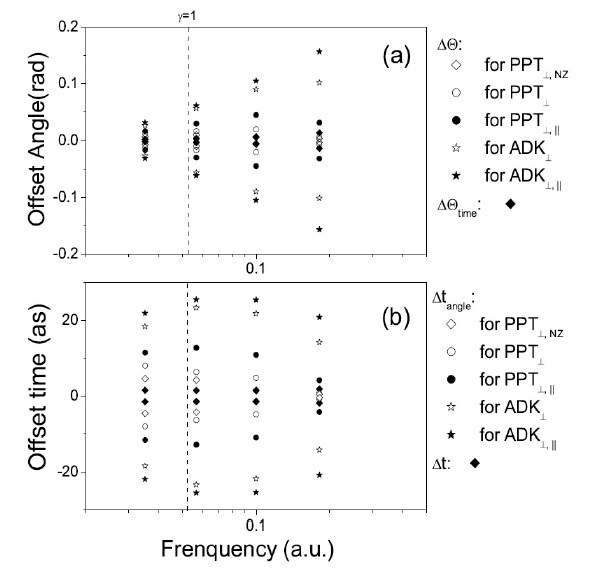}
\end{center}
\caption{ $\Delta\Theta$, $\Delta\Theta_{time}$ (a)
and $\Delta t$, $\Delta t_{angle}$ (b) obtained by the different
methods as shown in Fig. \ref{fig4} for four laser frequencies.
$\Delta\Theta_{time}$ ($\Delta t_{angle}$) means the angular
(ionization time) offset transformed from the offset time (angle)
(see text).
}%
\label{fig5}%
\end{figure}
It is well-known that the semiclassical model is only valid in the
quasistatic tunneling regime. Here, we also consider the
non-adiabatic effect using PPT theory \cite{popov,popov1}. The ionization rates for PPT are obtained by Eqs. (3.4)-(3.6) of \cite{popov}. The pre-exponential factor is taken from Eq. (59) of \cite{popov1}.
The initial momentum at the tunnel exit is
also taken into account in the same way as that in the
semiclassical model as mentioned above for an explicit comparison
with the semiclassical calculation. The PPT calculations including
the only initial transverse momentum and both the initial
transversal and longitudinal momenta at tunnel exit are denoted as
PPT$_{\perp}$ and PPT$_{\perp,\parallel}$, respectively. From
Figs. \ref{fig4}(a)-\ref{fig4}(c), we can find
that the peak width of PPT$_{\perp,\parallel}$ (dashed line) is
also broader than that obtained by PPT$_{\perp}$ (dashed dotted
dotted line) and the peak position of the latter one is closer to
the quantum peak position than the former one, which are
similar to the differences between ADK$_{\perp,\parallel}$ and
ADK$_{\perp}$ results. Furthermore, we also consider initial
transverse momentum with nonzero maximum in the PPT theory
(denoted by PPT$_{\perp,NZ}$ as shown in Figs. \ref{fig4}(a)-\ref{fig4}(c)).
The nonzero peak of the transverse momentum is given by
P$_{\perp}$=$\epsilon\gamma(t)\sqrt{2I_{p}}/6$ (here,
$\gamma(t)=\sqrt{2I_{p}}\omega/|E(t)|$ and $\epsilon$ is the
ellipticity) \cite{km}. It can be seen that the peak position is
closer to that of the quantum result than PPT$_{\perp}$ in the
angular distribution. Comparing the PPT calculations with the ADK
results, it can be seen that the peak positions calculated by PPT
are closer to those of the quantum theory, indicating that the
non-adiabatic effect can reduce the offset angle. It is worthwhile
mentioning that for the lower frequencies $\omega=0.05691$ and
0.03502 a.u., the widths of the angular distributions of PPT
theory are closer to the distributions of the quantum theory than
the results of semiclassical theory. However, for $\omega$=0.182
a.u., the angular distributions calculated by the PPT theory are
even broader than the distribution of the quantum theory, which
can also be seen clearly in Fig. \ref{fig4}(a). This indicates
that the PPT theory is invalid in the multiphoton regime. It is should be noted that for $\omega=0.05691$ in Fig. \ref{fig4}(b),
although there is a slight offset angle between the SFA and TDSE results due to a finite range of the potential used in the TDSE calculation, the value of offset angle is significantly smaller than any values of other offset angles under the same frequency condition. This implies that the calculation based on the SFA is accurate enough.

For the ionization time distribution ( see Fig. \ref{fig4}(d)),
the peak position of the semiclassical theory (black solid line)
corresponds to the maximal field strength. For the quantum
calculations, not only the peak width (shown clearly in Fig.
\ref{fig3} but also the peak position varies with
frequency, which becomes more and more close to those obtained by
the semiclassical theory (see Fig. \ref{fig4}(d)) when the
frequency decreases. For the PPT theory, although the peak widths
of the ionization time distributions for three frequencies are
different and become more and more broad with increasing
frequencies, the peak positions are the same as those of
the semiclassical theory and are independent of the laser
frequency, which means that the value of the offset time between
the quantum and ADK calculations is the same as that between the
quantum and PPT calculations for a given laser frequency. As shown
in Fig. \ref{fig4}(d), when the frequency decreases, the
ionization time distribution approaches the semiclassical
results, which can be attributed to that the non-adiabatic effect
is less and less important with decreasing Keldysh parameter.

Figure \ref{fig5} shows the values of the offset angles (Fig.
\ref{fig5}(a)) and offset times (Fig. \ref{fig5}(b)) extracted
from Fig. \ref{fig4}. The offset angles or times are positive and
negative, corresponding to the left and right peaks exhibited in
Figs. \ref{fig2} and \ref{fig3}, respectively. In order to
investigate the correspondence between the offset angle and the
offset time, we also give the corresponding offset angle
calculated from the offset time $\Delta t$ by the relation $\omega
t=\Theta$. Here we define this angular offset as
$\Delta\Theta_{time}=2\pi\Delta t$ ($\Delta t$ in units of
o.c.). Similarly, we define the corresponding offset time
transformed from the offset angle $\Delta\Theta$ in the angular
distribution as $\Delta t_{angle}$ ($\Delta
t_{angle}=\frac{\Delta\Theta}{2\pi}$). Fig. \ref{fig5}(a) shows
the offset angles $\Delta\Theta$ and $\Delta\Theta_{time}$ for
four different frequencies (the result of $\omega=0.1$ a.u. is
also shown here to exhibit the frequency dependence of the offset
time and angle more clearly). One can find that the offset angles
$\Delta\Theta$ of ADK$_{\perp}$, ADK$_{\perp,\parallel}$,
PPT$_{\perp}$ and PPT$_{\perp,\parallel}$ are always larger than
$\Delta\Theta_{time}$ for all frequencies except $\Delta\Theta$ of
PPT$_{\perp}$ for $\omega=0.182$ a.u. Moreover, the offset angle
of PPT$_{\perp,NZ}$, which is smaller than the other four offset
angles as mentioned before, is still larger than
$\Delta\Theta_{time}$ in the tunneling ($\gamma\ll$1) and
nonadiabatic regimes ($\gamma\approx$1) but becomes almost equal
to or even smaller than $\Delta\Theta_{time}$ in the multiphoton
regime.

\begin{figure}[tbh]
\begin{center}
\includegraphics[width=8.cm,height=7.cm]{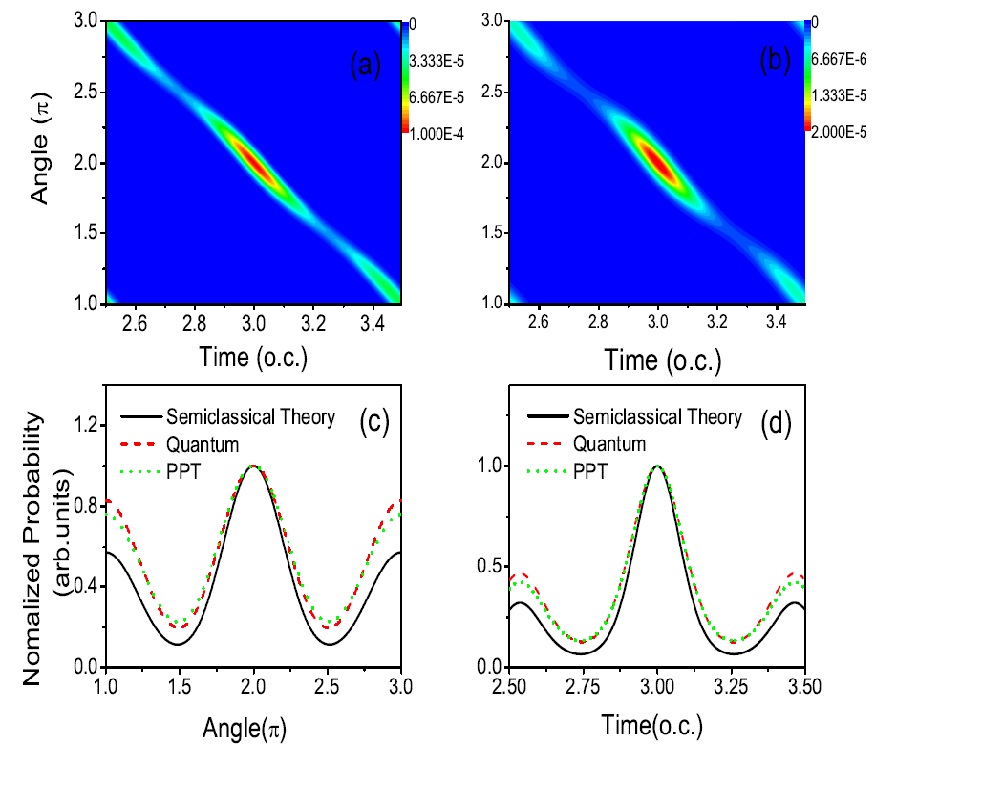}
\end{center}
\caption{Time-angle distributions calculated by
the quantum theory (a) and the semiclassical theory with
ADK$_{\perp,\parallel}$ (b) for $\omega$=0.05691 a.u.. Angular
distribution (c) and ionization time distribution (d) obtained by both
the quantum and semiclassical methods. The parameters are the same
as those of Fig. \ref{fig2}(c) except CEP $\varphi=0$.
}%
\label{fig6}%
\end{figure}
Figure. \ref{fig5}(b) depicts the offset time $\Delta t$ and $\Delta
t_{angle}$ in units of attosecond (as). Some interesting features
can be found in Fig. \ref{fig5}(b). The offset time hardly depends
on the frequency. When the frequency decreases from 0.182 a.u. to
0.03502 a.u., the offset time $\Delta t$ only changes from about
1.8 as to 1.5 as. However, the corresponding offset time $\Delta
t_{angle}$ varies in a much large range from 0.44 as to 26 as.
Generally, the offset time $\Delta t_{angle}$ of the PPT theory is
considerably smaller than that of the ADK theory, indicating that
the non-adiabatic effect will indeed reduce the discrepancy
between the quantum and semiclassical calculations. However, it
can be clearly seen that, except for the calculations of
PPT$_{\perp}$ and PPT$_{\perp,NZ}$ in the multiphoton regime
($\omega=$0.1 and 0.182 a.u.), the offset times $\Delta t$ are
noticeably smaller than the corresponding offset times $\Delta
t_{angle}$ transformed from the offset angle in the angular
distribution. This result implies that the offset angle
$\Delta\Theta$ in the angular distribution \emph{does not}
correspond to the offset time $\Delta t$ in the ionization time
distribution. So the "attoclock" measurement \cite{Eckle2} which
relies on the correspondence between the offset angle and time is
in principle inaccurate. It is noteworthy that there is a debate
on whether the non-adiabatic effect is responsible for the offset
angle measured for He in the experiment \cite{boge,Iva}. Our
calculation shows that although the non-adiabatic effect taken
into account in the PPT theory can reduce the value of the offset
angle, the offset angle between the quantum and PPT calculations
still does not correspond to the offset time between these two
calculations, which indicates that the discrepancy between the two
kinds of offsets cannot be attributed to the non-adiabatic effect
\cite{Yudin2001,Barth2011,Wang2014}.

\begin{figure}[tbh]
\begin{center}
\includegraphics[width=8.cm,height=6cm]{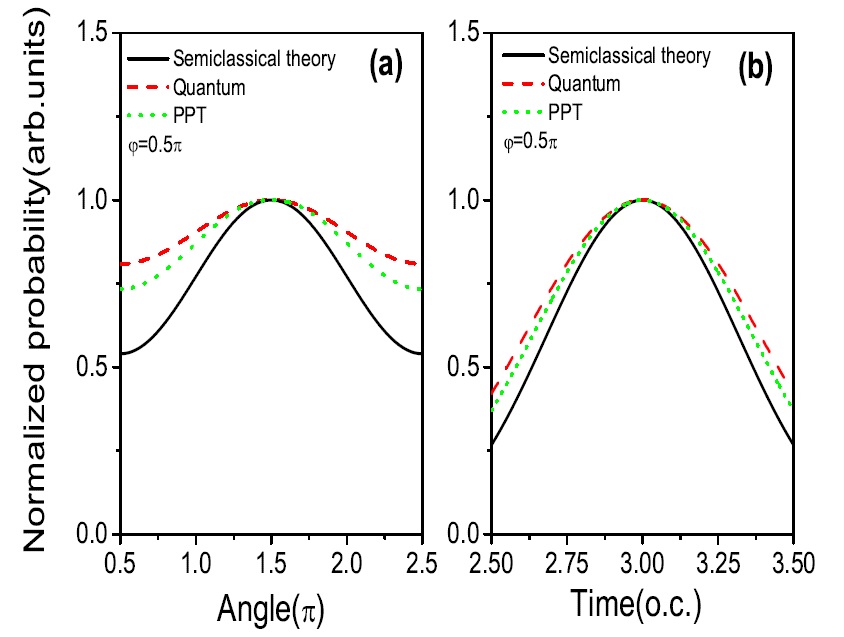}
\end{center}
\caption{ Angular (a) and ionization time (b)
distributions obtained by the quantum and semiclassical methods
for circularly polarized laser pulses. The other parameters are the same
as those of Fig. \ref{fig6}.
}%
\label{fig7}%
\end{figure}

It is worthwhile mentioning that for pulses with CEP
$\varphi=0.5\pi$, because the two peaks in the angular (ionization
time) distribution are symmetric, the absolute values of the
offset angles (times) for the two peaks are equal. For other CEPs,
the two peaks in the angular (ionization time) distribution become
asymmetric and the absolute values of the offsets for these two
peaks are not equal. Another special case is CEP $\varphi=0$ (see
Fig. \ref{fig6}). In this case, the maximal ionization rate occurs
at the center of the laser pulse at $t=3$ o.c. as shown in Fig.
\ref{fig6}. Figs. \ref{fig6}(a) and \ref{fig6}(b) are the
time-emission angle distributions calculated by the quantum theory
and the semiclassical theory with ADK$_{\perp,\parallel}$,
respectively. Both distributions show one main peak located at
[$\Theta,t]\sim[0(2\pi)$, 3 o.c.] and two other peaks at $t\sim$
2.536 and 3.464 o.c. Moveover, the angular distribution and the
ionization time distribution are given in Figs. \ref{fig6}(c) and
\ref{fig6}(d), respectively. The non-adiabatic effect is also
considered by PPT theory and the corresponding results are given
in Figs. \ref{fig6}(c) and \ref{fig6}(d). As displayed in Fig.
\ref{fig6}(c), all the angular distributions obtained by the
quantum theory, semiclassical theory and PPT theory show two
asymmetric peaks: one higher peak (namely one main peak) at
emission angle $\Theta=2\pi$ and one lower peak at emission angle
$\Theta=\pi(3\pi)$. The peak positions of the two peaks calculated
by these three methods coincide with each other, which means that
the offset angle for the both higher peak and lower peak are all
zero $\Delta\Theta=0$. However, the ionization time distributions
show a three-peak structure. For the highest peak at $t=3$ o.c.
which corresponds to the maximal field of the laser pulse, the
offset time $\Delta t$ is zero. For the other two peaks on the two
sides of the highest peak, these peaks are symmetric with respect
to the center of the pulse $t=3$ o.c. and the absolute values of
the offset times are equal $|\Delta t|=0.004$ o.c. (see Fig.
\ref{fig6}(d)). As shown in Figs. \ref{fig6}(c) and \ref{fig6}(d),
when the non-adiabatic effect is considered, the widths of both
the angular distribution and the ionization time distribution are
closer to the quantum distributions than the ADK results,
however, the positions of the peaks remain the same as those of
the ADK calculations. In addition, the calculations for circularly
polarized laser pulses are shown in Fig. \ref{fig7}. In this case,
there is only one peak in both the angular and ionization time
distributions and the peak positions both coincide with the ADK
(solid line) and PPT (dotted line) calculations, in agreement with
the results of Torlina \emph{et al.} \cite{Torlina2015} and Ni
\emph{et al.} \cite{nh}.

\subsection{Linear polarization}
\begin{figure}[tbh]
\begin{center}
\includegraphics[width=8.cm,height=6cm]{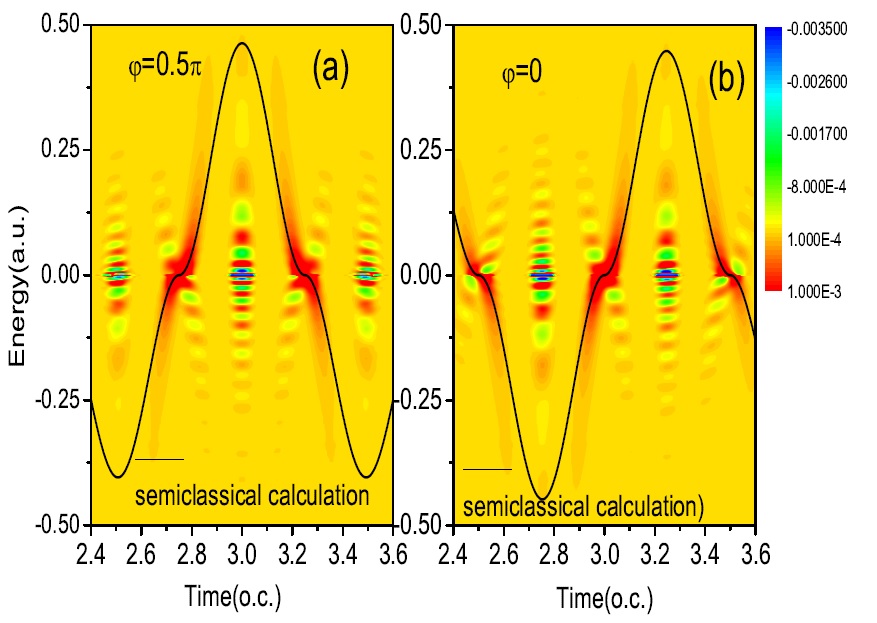}
\end{center}
\caption{ Time-energy distribution of photoelectron
in linearly polarized laser pulse. Here positive and negative
energies indicate the emission directions of the photoelectron. (a): CEP
$\varphi=0.5\pi$; (b): CEP $\varphi=0$. The other parameters are the
same as those of Fig. \ref{fig6}. The prediction of the simpleman's model is
also shown (see text).
}%
\label{fig8}%
\end{figure}
To systematically investigate the accuracy of the semiclassical
model, we also calculate photoionization process of atoms in
linearly polarized laser pulses. The time-energy distributions of
H atoms in an 800 nm 6-cycle laser field with peak intensity of
$1\times10^{14}$ W/cm$^2$ and CEPs of 0.5$\pi$ and 0 are depicted
in Fig. \ref{fig8}. For comparison, the prediction of the simpleman's model
is also shown in Fig. \ref{fig8}. In agreement with Fig. 2 of
\cite{pra-guo}, the correspondence between the instant of ionization
and the final kinetic energy (represented by strips in Fig. \ref{fig8}) is just
qualitatively consistent with the simpleman's prediction. Although the
maximum of the strips is close to the semiclassical
calculation near zero energy, two calculations diverge
with increasing energy still.

\begin{figure}[tbh]
\begin{center}
\includegraphics[width=8.cm,height=6.cm]{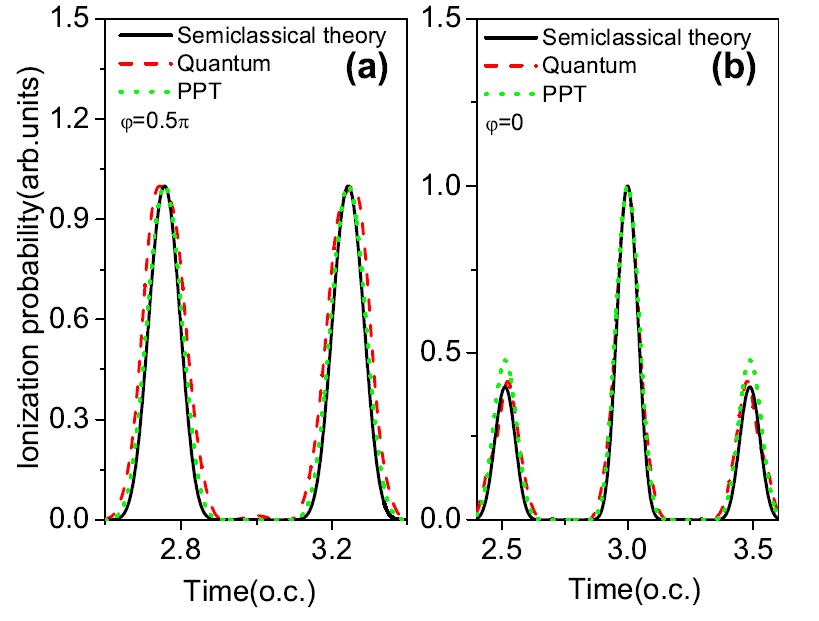}
\end{center}
\caption{ Ionization time distribution of
photoelectron in the linearly polarized laser pulses. (a): CEP
$\varphi=0.5\pi$; (b): CEP $\varphi=0$. The other parameters are
the same as those of Fig. \ref{fig6}.
}%
\label{fig9}%
\end{figure}

In Fig. \ref{fig9}, we show the ionization time distributions of
both quantum and semiclassical calculations. The results are
similar to the elliptical polarization case. For CEP
$\varphi=0.5\pi$, two peaks of both calculations are symmetric and
the offset times are also symmetric ($\Delta t=\pm32$ as). While
for $\varphi=0$, the higher central peaks of two calculations
coincide with each other but the lower side peaks show noticeable
offsets ($\Delta t=\pm29$ as). The PPT calculations are also given
in Fig. \ref{fig9} (dotted lines). The peak positions for PPT
theory are the same as those for semiclassical calculations and
the peak widths are broader than the semiclassical distribution,
which is similar to the case of the elliptical polarization. It
is noteworthy that the offset time in the linearly polarized case
is considerably larger than that in the elliptically polarized
laser field.

\subsection{Discussion}
\begin{figure}[tbh]
\begin{center}
\includegraphics[width=8.cm,height=5.5cm]{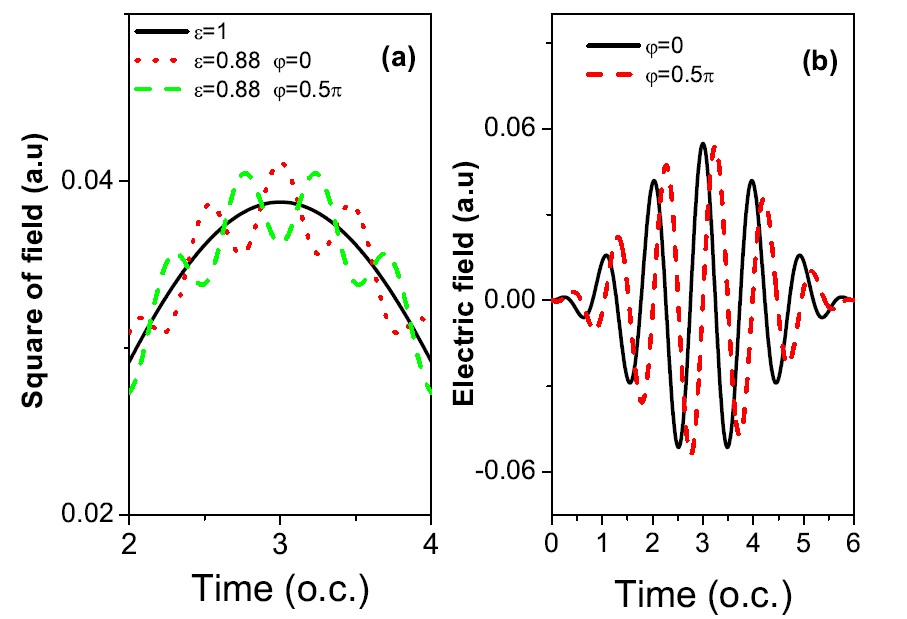}
\end{center}
\caption{(a): Square of electric field of
elliptically and circularly polarized laser pulses. (b): Electric
field of linearly polarized laser pulses for $\varphi=0$ and 0.5$\pi$ .
}%
\label{fig10}%
\end{figure}
In the Wigner-distribution-like function, to obtain the ionization
time distribution in the quantum theory, correlation effect must
be taken into account. The ionization probability at moment $t$ is
integration of the correlation between the ionization at $t+t'$
and $t-t'$ where $t'$ goes over the whole laser pulse, which means
that the quantum process is temporally nonlocalized. In contrast,
the ionization process in the semiclassical picture is temporally
localized, viz., ionization at one moment is independent of
ionization at other moment. Therefore, the semiclassical theory is
an approximate description of the photoionization process and, as
shown by our calculation, is usually not quantitatively consistent
with the quantum calculation. The coincidence of the two
calculations in some specific cases is actually accidental and can
be attributed to the time-reverse invariant symmetry of the laser
field. For an elliptically laser pulse with CEP $\varphi=0$ or a circularly
polarized laser pulse with any CEP, the laser field possesses
time-reverse invariant symmetry with maximum at the center of the
laser pulse as shown in Fig. \ref{fig10}. Therefore, the
ionization time and angular distributions are symmetrical and the
positions of the peaks (the higher peak in the elliptically
polarized case with $\varphi=0$), which are generated at the
maximum of the laser field, are independent of the calculation
method. For CEP $\varphi=0.5\pi$, though the laser field is also
time-reverse invariant, this symmetry cannot ensure the
coincidence between the peaks of different calculation methods but
guarantee that the offsets of the two symmetrical peaks in both
the ionization time and angular distributions are symmetrical
(positive and negative but the absolute values are the same, see.
Figs. \ref{fig2}- \ref{fig5}). Moreover, as mentioned before, the
lower peaks in the angular distribution for two calculations also
coincide with each other in the $\varphi=0$ case (see Fig.
\ref{fig6}(c)). This is because that the lower peak in the
angular distribution is actually composed of two halves
symmetrical with respect to $\Theta=\pi$ which are generated by
two symmetrical peaks with respect to $t$=3 o.c. in the ionization
time distribution (see Fig. \ref{fig6}(d)) and this symmetry is
independent of the calculation method. In the linearly polarized
cases, the offset time in the linearly polarized case is
considerably larger than that in the elliptically polarized laser
field. This can be attributed to stronger correlation effect in
the linearly polarized laser field since in this case, the emitted
photoelectrons concentrate more in one direction (laser field
direction) than that in the elliptically polarized field. For
other CEP, the time-reverse invariant is broken, hence the
distributions are asymmetrical and the offsets are also
asymmetrical. Moreover, our analysis indicates that neither the
offset time nor the offset angle in the elliptically polarized
laser field can be interpreted as the tunneling time delay. These
offsets, together with the widths of the peaks in the
distributions, however, could be taken as effective measures for
the validity of the semiclassical theory in description of the
ionization process in the laser field.

\section{Conclusion}

The WDL function enables us to obtain the time-emission angle
distribution, angular distribution and ionization time
distribution of atoms in the elliptically polarized laser fields and
time-energy distribution and ionization time distribution of atoms
in the linearly polarized laser fields. Comparison shows that these
distributions are usually not in quantitative agreement with the
semiclassical model calculation except in some specific cases.
Discrepancy between two calculations can be attributed to
correlation effect or temporal nonlocalization characteristic of
quantum mechanics and coincidence between two calculations in some
specific cases is due to temporal symmetry of the laser pulse.
Moreover, we find that the offset angle are generally not
consistent with the offset time even when the non-adiabatic effect
is taken into account, indicating that the attosecond angular
streaking technique is in principle inaccurate. Our result clearly
shows the applicability and limit of the technique, which is
important for interpretation and understanding on relevant
experimental and theoretical results.

\section*{Funding}
This work was partially supported by the National Key program for
S\&T Research and Development (No. 2016YFA0401100), NNSFC (No.
11774361, 11775286, 11425414 and 11774215)and the Open Fund of the State Key Laboratory of High Field Laser Physics (SIOM).


\end{document}